\documentclass[prl,twocolumn,amsmath,amssymb]{revtex4}
\usepackage{graphicx}
\usepackage{bm}
\usepackage{amsmath,amssymb}
\usepackage{color}
\usepackage{ascmac}

\begin{document}
\title{Highly-mixed measurement-based quantum computing and the one clean qubit model}  
\author{Tomoyuki Morimae}
\email{morimae@gunma-u.ac.jp}
\affiliation{ASRLD Unit, Gunma University,
1-5-1 Tenjin-cho Kiryu-shi Gunma-ken, 376-0052, Japan}

\date{\today}
            
\begin{abstract}
We show that a highly-mixed state in terms of a large
min-entropy is useless as a resource state 
for measurement-based quantum computation
in the sense that if a classically efficiently verifiable
problem is efficiently solved   
with such a highly-mixed measurement-based quantum computation
then such a problem can also be classically efficiently solved.
We derive a similar result also for
the DQC1$_k$ model,
which is a generalized version of
the DQC1 model where $k$ output qubits are measured.
We also show that the measurement-based quantum computing
on a highly-mixed resource state
in terms of the von Neumann entropy,
and DQC1$_k$ model are useless in another sense that
the mutual information between the computation results and
inputs is very small.
\end{abstract}

\maketitle  
One of the most fundamental questions in quantum information
science is whether a quantum computing model truly outperforms 
classical computing or not.
In particular, to clarify the power of quantum computing models that are highly mixed
is important both from the fundamental and practical points of view.
In this paper, we consider two highly-mixed quantum computing
models, namely, the measurement-based quantum computing 
on a highly-mixed resource state, and the one clean qubit model with many-qubit measurements.
We show that if a classically efficiently verifiable
problem is efficiently solved with these models, then such a problem can also be
classically efficiently solved.
In this sense,
these two models are not useful.
We also show that for these highly-mixed quantum computing models,
the mutual information between the computation outputs and inputs is very small.
This means that these models are useless for problems where inputs and outputs should
be highly correlated like a search problem.

Measurement-based quantum computation (MBQC) by Raussendorf and Briegel~\cite{MBQC}
is a model of quantum computing where
universal quantum computation can be done
with only local measurements on a certain quantum many-body
state, which is called a resource state,
and a classical processing of the measurement results.
The computational power of MBQC is equivalent to the traditional circuit
model of quantum computation, but the clear separation between the 
quantum phase (i.e., the preparation of the resource state) 
and the classical phase (i.e., local adaptive
measurements) has inspired many new results over the last decade,
which would not be obtained from the circuit model mind.
For example, new resource states for MBQC which are closely connected with
condensed matter physics have been proposed~\cite{Verstraete,Gross_QCTN,
MiyakeAKLT,Cai,Miyake_edge,Miyake2dAKLT,Wei2dAKLT,Cai_magnet,fMBQC,
upload,stringnet}. 
Furthermore, relations between MBQC and partition functions
of classical spin models were pointed out~\cite{Bravyi,Nest,Nest2}.
These discoveries have established new bridges between
quantum information and condensed matter physics.
MBQC has also offered a new framework of fault-tolerant
quantum computing, namely, the topological measurement-based
quantum computation, which achieves dramatically high 
error thresholds~\cite{Raussendorf_topo,
Sean,FujiiTokunaga,Ying,Ying2,FMspin}.
New protocols of secure cloud quantum computing,
so called the blind quantum computing, were also developed
by using MBQC~\cite{BFK,FK,Barz,Vedran,
AKLTblind,topoblind,CVblind,topoveri,MABQC,Sueki,composable,
composableMA,distillation,Lorenzo,Joe_intern,Barz2,honesty}. 

MBQC on mixed resource states have been studied 
by several researchers~\cite{RBH,Ying2,FMspin,Kwek,BBDJR}.
In Refs.~\cite{Ying2,FMspin}, 
some condensed-matter physically motivated two-body Hamiltonians were proposed
whose equilibrium states at sufficiently low
temperatures can be used as resource states for the
topologically-protected MBQC.
In Ref.~\cite{RBH}, the thermal three-dimensional
cluster state was considered, 
and it was shown that the entanglement length can be infinite 
if the temperature is below a certain threshold,
whereas it becomes finite if the temperature is higher than
another certain threshold. 
In Ref.~\cite{Kwek}, a two-body qubit Hamiltonian was proposed
whose low-temperature equilibrium states
can be adiabatically brought to useful resource states
for the topologically-protected MBQC.
In Ref.~\cite{BBDJR},
thermal cluster states are considered, and it was shown that
at a low-temperature region these thermal states can be universal
resources whereas at a high-temperature region MBQC on these
thermal states can be classically efficiently simulated.
However, all these results consider only equilibrium states,
or assume specific forms of Hamiltonians or resource states.

In this paper, 
we obtain a general result that highly mixed states in terms of a 
large min-entropy are useless resource states for MBQC in the sense that
if a classically efficiently verifiable problem
is efficiently solved with MBQC on such states,
then such a problem can also be classically efficiently solved.
The result is general: we do not make any assumption on Hamiltonians, 
resource states,
or the way of measurements, etc.
The min-entropy $H_{min}(\rho)$ 
$(0\le H_{min}(\rho)\le N)$ of an $N$-qubit
state $\rho$ is defined by
$H_{min}(\rho)\equiv-\log_2\lambda_1$,
where $\lambda_1$ is the largest eigenvalue of $\rho$.
The min-entropy quantifies the amount of random bits that can be extracted~\cite{Konig}.
Our result is derived by using a similar argument of Ref.~\cite{Gross_ent}
that shows that highly-entangled pure states in terms of the geometric
measure of entanglement~\cite{GM1,GM2,GM3} are useless resource states
for MBQC in a similar sense.
As in Ref.~\cite{Gross_ent}, we first assume that a quantum computing model can 
efficiently solve a problem whose solution can be classically efficiently
verifiable. Then we show that we can construct
a classical random computing model that can efficiently solve the same problem.
Note that the uselessness of a randomly chosen pure state
as a resource state of MBQC
was shown in Ref.~\cite{Bremner}.

In particular, our result implies that the equilibrium state,
$e^{-\beta H}/\mbox{Tr}(e^{-\beta H})$, of any Hamiltonian $H$
with a high temperature $(\beta\ll1)$ are useless resource states for MBQC, 
where $\beta\equiv1/(kT)$, $k$ is the Boltzmann constant,
and $T$ is the temperature.
It is a generalization of the above mentioned previous results that assume specific Hamiltonians 
or resource states to any Hamiltonian and resource state.
Furthermore, we can also obtain the following result:
in order to change such a useless resource state into a useful one, 
$O(N)kT$ of work is necessary for the isothermal process,
where $N$ is the number of particles of the resource state. 

\begin{figure}[htbp]
\begin{center}
\includegraphics[width=0.4\textwidth]{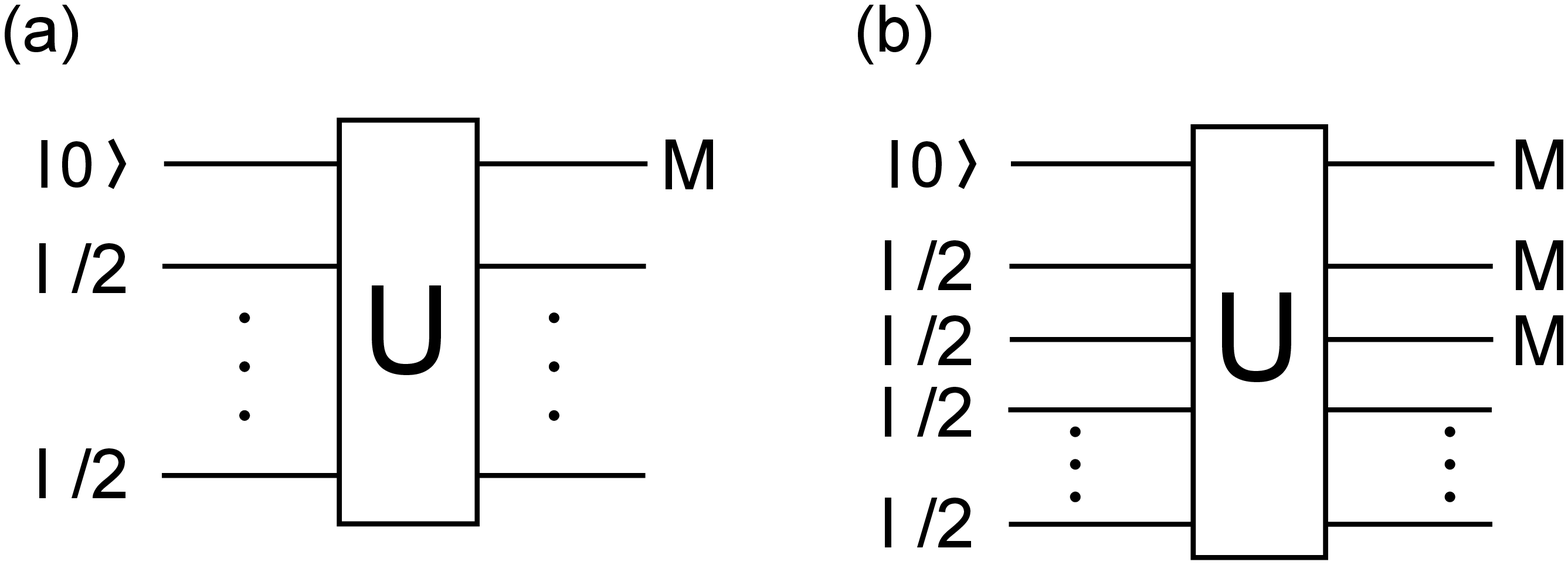}
\end{center}
\caption{
(a) the DQC1 model. (b) the DQC1$_k$ model for $k=3$.
Here, $U$ is an $n+1$ qubit unitary gate,
and $M$ is the computational basis measurement.
} 
\label{DQC1}
\end{figure}

We also consider  
another model of highly-mixed quantum computing, namely, the DQC1$_k$ model,
and derive a similar result:
if a classically efficiently verifiable problem is 
efficiently solved with the DQC1$_k$ model
then such a problem can also be classically efficiently solved.
Here, the DQC1$_k$ model is a generalized version
of the deterministic quantum computation with one quantum bit
(DQC1) model by Knill and Laflamme~\cite{KL}.
As is shown in Fig.~\ref{DQC1} (a), a DQC1 circuit consists of 
the input state, 
$\rho_{in}\equiv|0\rangle\langle0|\otimes \left(\frac{I}{2}\right)^{\otimes n}$, 
where $I\equiv|0\rangle\langle0|+|1\rangle\langle1|$ is the two dimensional identity operator, 
polynomial number of quantum gates on it,
and the computational basis measurement of the first qubit. 
The DQC1$_k$ model is equivalent to the DQC1 model except that not the single
but $k$ output qubits are measured in the computational basis 
at the end of the computation (Fig.~\ref{DQC1} (b)).
Surprisingly, such highly-mixed quantum computing models
can efficiently solve some problems for which no efficient classical 
algorithms are known, such as
the spectral density estimation~\cite{KL}, testing integrability~\cite{Poulin}, 
calculation of the fidelity decay~\cite{Poulin2},
and approximations of the Jones polynomials, the HOMFLY polynomials,
and the Turaev-Viro invariant~\cite{SS,Passante,JW,Alagic}.
Furthermore, it was shown in Ref.~\cite{MFF} that if any output probability distribution of
DQC1$_k$ for $k\ge3$ can be classically efficiently sampled within a certain multiplicative error,
the polynomial hierarchy~\cite{Toda} collapses at the third level, which is
not believed to happen.

We show that a classically efficiently verifiable problem
which can be efficiently solved with the DQC1$_k$ model
can also be classically efficiently solved.
Note that there is another negative result about the power of the DQC1 model:
the DQC1 model cannot simulate universal quantum computation
under some reasonable assumptions~\cite{Ambainis}.

Finally, 
we further show another negative results that MBQC
on a highly-mixed resource state in terms of the von Neumann entropy,
and the DQC1$_k$ model are useless in another sense that
the mutual information between the computation results and
inputs is very small. This means that these models are not useful
for problems where inputs and outputs should be strongly correlated like
search problems.

\begin{figure}[htbp]
\begin{center}
\includegraphics[width=0.2\textwidth]{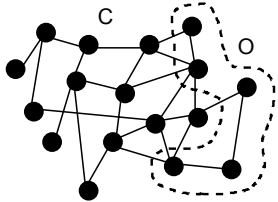}
\end{center}
\caption{
The resource state $\sigma$.
Filled circles are qubits.
The subsystem $O$ is the set of qubits in the region
indicated by the dotted line.
The subsystem $C$ is the set of qubits which are not in $O$.
} 
\label{figMBQC}
\end{figure}

{\it MBQC}.---
Before giving our first result about MBQC, let us define
the most general framework of MBQC.
Let $\sigma$ be the $N$-qubit resource state of MBQC.
(We can also consider qudit states for $d\ge3$, but for the simplicity,
we here consider qubit states. Generalizations to qudit states with
$d\ge3$ are straightforward.)
Note that $\sigma$ is not necessarily the graph state.
It can be any resource state, such as the Affleck-Kennedy-Lieb-Tasaki (AKLT) 
state~\cite{AKLT,MiyakeAKLT,
Cai,Miyake_edge,Miyake2dAKLT,Wei2dAKLT,Cai_magnet}
or a general tensor-network state~\cite{Gross_QCTN}.
As is shown in Fig.~\ref{figMBQC},
the resource state $\sigma$ is divided into two subsystems, $C$ and $O$,
which consists of $N-n$ and $n$ qubits, respectively.
Qubits in the subsystem $C$ are measured to perform the desired
quantum computation.
The input state of the computation is included in $C$.
We perform
a POVM $\{M_j\}_{j=1}^r$ on $C$ and obtain the result 
$m\in\{1,2,...,r\}$,
where $\sum_{j=1}^rM_j=I^{\otimes N-n}$.
(Local adaptive projective measurement used in usual MBQC~\cite{MBQC}
is a special case of the POVM. Here, we consider the most general way of MBQC,
hence we use POVM, which can be global.)
After the POVM,
the output of the computation is encoded on qubits in $O$. 
We measure the subsystem $O$ in the computational basis
in order to read out the output of the quantum computation.
Let us denote the computational basis measurement
on $O$ by $\{P_z\}$,
where $z\in \{0,1\}^n$
is an $n$ bit binary string, and
$P_z\equiv|z\rangle\langle z|\equiv\bigotimes_{j=1}^n|z_j\rangle\langle z_j|$
is the projection operator onto the computational basis
$|z\rangle\equiv\bigotimes_{j=1}^n|z_j\rangle$.
Here, $z_j\in\{0,1\}$ is the $j$th bit of $z$.
Depending on the outcome $m\in\{1,2,...,r\}$ 
of the previous POVM $\{M_j\}_{j=1}^r$ on $C$, 
the result $z$ of the computational basis measurement on $O$ 
is classically post-processed in order to correct the
effect of the byproduct operators, which are unavoidable
in MBQC~\cite{MBQC,byproduct}.

{\it Highly mixed states are useless resource states.}---
Now let us show our first result that 
highly mixed states are useless resource states for MBQC.
As in Ref.~\cite{Gross_ent}, we consider 
MBQC solving a classically efficiently verifiable problem, and require that the probability
of obtaining a correct result is larger than $1/2$~\cite{path}:
\begin{eqnarray}
\frac{1}{2}\le
\sum_{j=1}^r\sum_{z\in S_j}
\mbox{Tr}[(M_j\otimes P_z)\sigma],
\label{prob}
\end{eqnarray}
where $S_j\subseteq \{0,1\}^n$ is the set of correct results when 
the result of the POVM on $C$ is $j$.
(Note that the set of correct results depends on the result of
the previous POVM on $C$ because of the byproduct operators.)
Because we assume that our MBQC is deterministic, 
i.e., any byproduct operators are correctable,
we require that
$|S_1|=|S_2|=...=|S_r|=|S|$.
(It is interesting to consider non-deterministic MBQC,
but it would be a subject of a future study.)
Let 
$\sigma=\sum_{k=1}^{2^N}\lambda_k|\lambda_k\rangle\langle \lambda_k|$
be the spectral decomposition (i.e., the diagonalization) 
of the resource state $\sigma$ 
with the decreasing order 
$1\ge\lambda_1\ge\lambda_2\ge...\ge\lambda_{2^N-1}\ge\lambda_{2^N}\ge0$ 
of eigenvalues and corresponding eigenvectors 
$\{|\lambda_k\rangle\}_{k=1}^{2^N}$.
Then, Eq.~(\ref{prob}) becomes
\begin{eqnarray*}
\frac{1}{2}
&\le&\sum_{j=1}^r\sum_{z\in S_j}\sum_{k=1}^{2^N}
\lambda_k
\mbox{Tr}\Big[(M_j\otimes P_z)|\lambda_k\rangle\langle\lambda_k|\Big]\\
&\le&\lambda_1|S|2^{N-n}.
\end{eqnarray*}
This means 
that
$|S|\ge2^{-N+n-1}\lambda_1^{-1}$.
Let us assume that we randomly generate an $n$-bit binary string
$a\in\{0,1\}^n$.
Since the problem is classically efficiently verifiable, we can efficiently check whether
the string $a$ is a solution or not.
If it is not a solution, we again generate another
random $n$-bit binary string $a'\in\{0,1\}^n$, and check whether
it is a solution or not. We repeat this process 
until we finally obtain a correct solution.
Let us assume that $H_{min}(\sigma)=N-\delta$,
where $\delta$ is at most $O(\log N)$.
This means that $\lambda_1=2^{-N+\delta}$.
Now we can use the result of Ref.~\cite{Gross_ent}:
the probability that we do not obtain any correct string
after $t$ repetitions of the above process is 
$\Big(1-\frac{|S|}{2^n}\Big)^t
\le(1-2^{-N-1}\lambda_1^{-1})^t
=(1-2^{-\delta-1})^t
<e^{-t2^{-\delta-1}}$.
If we take $t=2^{\delta+1}\ln(1/p_f)$,
the probability of failure is less than $p_f$.
Therefore, a correct result can be efficiently classically obtained
with a sufficiently large success probability.

{\it Example 1: equilibrium states}.---
As an example, let us consider the special case that
the resource state $\sigma$ is the equilibrium state, 
$\sigma=e^{-\beta H}/\mbox{Tr}(e^{-\beta H})$,
of a Hamiltonian $H$,
where $\beta=(kT)^{-1}$.
Let us scale the energy spectrum of the Hamiltonian so that
the lowest energy of $H$ is 0.
Then, the min-entropy of $\sigma$ is given by 
$H_{min}(\sigma)
=\log\mbox{Tr}(e^{-\beta H})=-\beta F$,
where $F$ is the free energy.
Let us assume $H_{min}(\sigma)=N-\delta$,
where $\delta$ is at most $O(\log N)$.
In other words, we assume that $\sigma$ is a useless resource state.
If we want to change $\sigma$ into another state $\sigma'$
that is useful, 
$H(\sigma')$ must be $N-\delta'$, 
where $\delta'=O(N)$.
Then,
$H_{min}(\sigma)-H_{min}(\sigma')=\delta'-\delta=O(N)$.
This means that the free energy change is $\Delta F=O(N)kT$,
and therefore $O(N)kT$ of work is necessary for the isothermal
transformation.

{\it Example 2: thermal cluster state}.---
Our next example is the $N$-qubit thermal cluster state
$\sigma_{cl}\equiv e^{-\beta H_{cl}}/\mbox{Tr}(e^{-\beta H_{cl}})$,
where
$H_{cl}\equiv-\sum_{j=1}^NK_j$ is the cluster state Hamiltonian,
$K_i\equiv X_i\bigotimes_{j\in N(i)}Z_j$ is a stabilizer
operator for the cluster state.
Here, $X_j$ and $Z_j$ are Pauli $X$ and $Z$ operators acting on site $j$,
and $N(i)$ is the set of the nearest-neighbour sites of site $i$.
Note that $[K_i,K_j]=0$ for all $i$ and $j$.
Let $|C\rangle$ be the ground state 
(i.e., the cluster state) of the Hamiltonian $H_{cl}$:
$H_{cl}|C\rangle=-N|C\rangle$.
Then, the set of states
$|s\rangle\equiv\Big(\bigotimes_{j=1}^NZ_j^{s_j}\Big)|C\rangle$,
for $s\equiv(s_1,...,s_N)\in\{0,1\}^N$
is an orthonormal basis of the $2^N$-dimensional Hilbert space.
Therefore, we can easily calculate $H_{min}(\sigma_{cl})$ as
$H_{min}(\sigma_{cl})=
-\log_2\frac{e^{\beta N}}{\mbox{Tr}(\prod_{i=1}^Ne^{-\beta K_i})}
=N\log_2(1+e^{-2\beta})$.
If $T\to\infty$, then $\beta\to0$, and therefore
$H_{min}(\sigma_{cl})\to N$. 
This means that the thermal cluster state
with a high temperature is useless resource state
for MBQC.

{\it DQC1$_k$ model.}---
Now let us show our second result about the DQC1$_k$ model.
We consider the DQC1$_k$ model
solving a classically efficiently verifiable problem. We again require that the probability
of obtaining a correct result is larger than $1/2$:
\begin{eqnarray}
\frac{1}{2}\le
\sum_{z\in S}
\mbox{Tr}\Big[(P_z\otimes I^{\otimes n+1-k})U\rho_{in}U^\dagger\Big],
\label{prob2}
\end{eqnarray}
where $P_z\equiv|z\rangle\langle z|$ is the $k$ qubit projection
operator onto the computational basis, $z\in\{0,1\}^k$ is a $k$ bit string,
and $S\subseteq \{0,1\}^k$ is the set of correct results.
Let 
$U\rho_{in}U^\dagger=\sum_{j=1}^{2^{n+1}}\lambda_j|\lambda_j\rangle\langle \lambda_j|$
be the spectral decomposition (i.e., the diagonalization) 
of $U\rho_{in}U^\dagger$ 
with the decreasing order 
$1\ge\lambda_1\ge\lambda_2\ge...\ge\lambda_{2^{n+1}-1}\ge\lambda_{2^{n+1}}\ge0$
of eigenvalues and corresponding eigenvectors 
$\{|\lambda_j\rangle\}_{j=1}^{2^{n+1}}$.
Since the unitary $U$ does not change the spectrum of $\rho_{in}$,
\begin{eqnarray*}
\lambda_j=
\left\{
\begin{array}{ll}
2^{-n}& (1\le j\le 2^n)\\
0&(2^n+1\le j\le 2^{n+1}).
\end{array}
\right.
\end{eqnarray*}
Therefore, Eq.~(\ref{prob2}) becomes
\begin{eqnarray*}
\frac{1}{2}
&\le& 
\sum_{z\in S}\sum_{j=1}^{2^{n+1}}\lambda_j
\mbox{Tr}\Big[(P_z\otimes I^{\otimes n+1-k})
|\lambda_j\rangle\langle\lambda_j|\Big]\\
&\le& 
\lambda_1|S|2^{n+1-k}.
\end{eqnarray*}
This means that 
$|S|\ge\lambda_1^{-1}2^{-n+k-2}=2^{k-2}$.
Let us generate a random $k$ bit string and check whether it is a solution 
or not. Since the problem is classically efficiently verifiable, we can check efficiently.
If it is not a solution, we again generate another random $k$ bit string. 
The probability that we do not obtain any correct solution after
$t$ repetition of the process is
$\Big(1-\frac{|S|}{2^k}\Big)^t\le \Big(1-\frac{1}{4}\Big)^t
=\Big(\frac{3}{4}\Big)^t$.
Therefore, $t=poly(n)$ of the repetition is sufficient to obtain a correct solution
with an exponentially small failure probability.
In short, a classically efficiently verifiable
problem which can be efficiently solvable with the DQC1$_k$ model
can also be classically efficiently solved.

Finally, we further show that
MBQC on a highly-mixed resource state
in terms of the von Neumann entropy, and DQC1$_k$ model 
are useless in another sense that the correlation between computation outputs and inputs 
quantified by the mutual information
is very small.
Let us consider the bipartite quantum computing between Alice and Bob:
Alice has the input, and she asks Bob to perform quantum computing.
After the computation, Bob sends Alice the output of the computing. 
Alice's classical input $i$ is represented by an $N_A$-qubit
state $|i\rangle$. For example, if Bob does MBQC, $i$ is the instruction of
how to measure each qubit, i.e., the measurement angles and the
way of adaptation, etc.
If Bob does the DQC1$_k$ computing, $i$ specifies the unitary that should be implemented,
etc.
Since different classical inputs $i$ and $i'$ must be distinguishable,
we assume that $\{|i\rangle\}_{i=1}^{2^{N_A}}$ is an orthonormal basis.
The initial state of the bipartite computing
is 
\begin{eqnarray*}
\Big(\sum_{i=1}^{2^{N_A}}p_i|i\rangle\langle i|_A\Big)\otimes \sigma_B,
\end{eqnarray*}
where the classical input $|i\rangle$ is generated with the probability $p_i$,
and $\sigma_B$ is Bob's initial quantum state of $N_B$ qubits.
Bob applies the unitary $U_i$ if Alice's input is $i$.
Then the output state of the bipartite computing is
\begin{eqnarray*}
\rho_{AB}=\sum_{i=1}^{2^{N_A}}p_i|i\rangle\langle i|_A\otimes U_i\sigma_BU_i^\dagger.
\end{eqnarray*}
(Note that if Bob does MBQC, he further measures his part of this state.
Since we are deriving a negative result, it is sufficient to consider the
positive branch, i.e., the case when accidentally no feedfowarding was required.)
The mutual information between $A$ and $B$ of $\rho_{AB}$ is
$I_{AB}=S(\rho_B)-\sum_{i=1}^{2^{N_A}}p_iS(\sigma_B^i)
\le N_B-S(\sigma_B)$,
where $\rho_B$ is the reduced density operator of $\rho_{AB}$
for Bob's system, and $\sigma_B^i=U_i\sigma_BU_i^\dagger$.
If $S(\sigma_B)=N_B-\delta$, for a certain small $\delta$,
then $I_{AB}\le \delta$,
which means that the mutual information is bounded by $\delta$.
For example, if Bob does MBQC, $\sigma_B$ is an $N$-qubit resource state,
i.e., $N_B=N$, and the unitary $U_i$ is the rotation of each qubit according 
to the specification $i$.
If $\delta$ is very small, the output of the MBQC is not sufficiently correlated
with Alice's inputs, and therefore in that sense such an MBQC is useless.
On the other hand, if Bob does the DQC1$_k$ computing, $\sigma_B$ is the highly-mixed input state 
$\rho_{in}=|0\rangle\langle0|\otimes (\frac{I}{2})^{\otimes n}$
of DQC1$_k$ model (and therefore $N_B=n+1$).
In the latter case, $S(\sigma_B)=n$, and therefore $I_{AB}\le 1$.

The author thanks Keisuke Fujii for valuable discussion.
This work was supported by the Tenure Track System MEXT Japan,

{\it Appendix}.---
We here show that parallelization of DQC1$_k$ circuits does not
change the result.
We allocate $r$ DQC1$_k$ circuits in parallel.
Let $o_i^j\in\{0,1\}^k$ be $i$th outcome of $j$th DQC1$_k$ circuit,
where $i=1,2,...,2^k$ and $j=1,2,...,r$.
We can repeat such a parallel computing in $v$ times.
We assume that some sets 
$\{(o_{f(1,j)}^1,o_{f(2,j)}^2,o_{f(3,j)}^3,...,o_{f(r,j)}^r)\}_j$
are solutions of our problem.
If we denote
$p_i=\sum_{z\in S_i}\mbox{Tr}[(P_z\otimes I^{n+1-k})\sigma]$,
where 
$S_i=\{o_{f(i,j)}^i\}_j$,
then $p_iv$ must be increasing as a function of $v$,
since
the probability that we never obtain any element of $S_i$ in the $v$ time repetition
is
$(1-p_i)^v\to e^{-p_iv}$.
Then, we obtain 
$p_i\le |S_i|\lambda_12^{n+1-k}$,
which means 
that the failure probability of the classical sampling is
$p_f= (1-\frac{|S_i|}{2^k})^v
\le (1-\frac{p_i}{2})^v
\le e^{-\frac{p_iv}{2}}
\to0$.



\begin{thebibliography}{99}
\bibitem{path}
Note that here we consider the classical simulation of only the output port (i.e.,
the subsystem $O$).
In Ref.~\cite{Gross_ent}, the classical simulation of all measurement histories,
i.e., measurement results on both $C$ and $O$,
is considered.
Our result is not changed if we consider such a case, since
$\frac{1}{2}\le\sum_{z\in S}\mbox{Tr}(P_z\sigma)$
leads to $\frac{|S|}{2^N}\ge \lambda_1^{-1}2^{-1}$,
where $S\subseteq\{0,1\}^N$ is the set of ``correct paths".
\end{thebibliography}
\end{document}